latex file plus three postscript files.

\documentstyle[preprint,aps]{revtex}
\begin{document}

\draft

\title{Quark Pair Production In The Chiral Phase Transition}

\author{Carsten Greiner}

\address{Department of Physics, Duke University, Durham NC 27708-0305}

\date{\today}

\maketitle

\begin{abstract}
The production of quarks and antiquarks during a sudden
restoration of chiral symmetry, as it might occur
in very energetic heavy ion collisions, is considered.
If gluons are already present they can assist additively
to the overall production:
real gluons are partially decaying during such a chiral transition.
The total number of produced quarks is calculated and found to be
quite sizeable. It is speculated that such a phenomenon could
give rise to a significant
contribution to the overall quark pair creation
in the preequilibrium stage of the heavy ion collision.
\end{abstract}

\pacs{25.75.+r, 11.10.Qr, 12.38.Bx, 24.85.+p}
\bigskip\medskip

One primary goal in the upcoming relativistic heavy ion collision
experiments at the Brookhaven Relativistic Heavy Ion Collider (RHIC)
and the CERN Large Hadron Collider (LHC) is the
temporary formation and potential
observation of a new form of matter, the {\em quark gluon plasma}, a
deconfined phase of QCD \cite{QM}. It is expected that this phase is also
accompanied by a restoration of chiral symmetry \cite{CS}.
At high enough temperatures the massive
and confined quasiparticles, the constituent quarks, become bare, undressed
quarks with current quark masses being much lighter than the constituent
quark masses. Such a physical picture is realized
in model descriptions of low-energy, effective
QCD lagrangians like the Nambu and Jona-Lasinio (NJL) model \cite{NJL,Sch92}:
At large enough coupling chiral symmetry of the vacuum
is spontaneously broken and exhibits a nonvanishing scalar quark condensate
$\langle \bar{q} q \rangle \, \simeq \, -(250\, \mbox{MeV})^3$.
A similar behaviour
is experienced in lattice QCD calculations \cite{Go89}.
The scalar condensate attributes
a scalar selfinteraction to the fermion propagator resulting in
the presence of constituent quarks at low energies, in line
with experience
from hadronic spectroscopy. For higher temperatures, however, the condensate
becomes weaker and vanishes at sufficiently large temperatures (on the order of
$T \sim 200 $ MeV)
\cite{GL89,NJL}, which
is also found in recent QCD calculations \cite{Ko91}.

Relativistic heavy ion collisions offer the only possible way to gain
more insight into the phase structure of nuclear matter at such high
temperatures
(and densities). However, from the onset the reaction during any such a
collision
proceeds far off equilibrium, so that a thermal quark gluon plasma may be
reached
only after some finite time.
According to the results of the parton cascade model \cite{Ge1} the quarks
will reach, if at all, a chemical saturation only eventually
in a central relativistic heavy ion collision at collider energies whereas
the gluons saturate on a significantly smaller time scale (in about
half a fm/c) \cite{Shu92,Ge2}. This `hot glue' scenario was proposed
by Shuryak \cite{Shu92}. The reason for
this particular behaviour lies in the fact that the quarks are only produced in
second order by gluon annihilation and the rate for this process is small
compared to the corresponding one for gluon production.  The decay of a
single gluon into a quark and antiquark is, of course, kinematically
not allowed, as long as all partons are treated as effective on-shell
particles.
In this work we want to propose that the occurrence
of a chiral phase transition can have some profound influence on the
quark pair production, if the restoration will happen on a very short
timescale. The idea is that if indeed a deconfined state
(equilibrated or not) will
be formed in the reaction of a heavy ion collision, the `undressing' of
the quarks should already have occurred. Being probably still
far from equilibrium this chiral restoration may thus happen rather
spontaneously. We assume that the scalar quark condensate
$\sigma \sim \langle \bar{q} q \rangle$ would melt on a timescale smaller than
any
scale set by the dynamics. The decay of the vacuum induces a spontaneous
creation of quark and antiquark pairs.
During the chiral restoration transition the quarks are no longer
on-shell particles, their spectral function contains a wide off-shell spectrum,
thus permitting also the decay of a single gluon
into a quark-antiquark pair. The decay of both the vacuum and a substantial
fraction of energetic gluons would accelerate the chemical saturation of the
quark phase space.

To describe the effect on quark pair production
we construct a simple model for fermionic matter.
Denoting the time of chiral restoration at $t=0$, the
light quark masses then are functions of time
\begin{equation}
\label{1}
m(t) \, = \, \left\{ \begin{array}{rrrr} m_c &\sim &350 \, \mbox{MeV}
& \mbox{for} \, t \leq 0 \\
m_b &\sim &10 \, \mbox{MeV} & \mbox{for} \, t>0 \end{array} \right. \, \, .
\end{equation}
We assume that no further residual interactions act among the quarks.
While ansatz (\ref{1}) may appear quite artificial, it captures
the essential spirit of what one would naively call `a transition to the
phase with restored chiral symmetry'.
The constituent vacuum becomes unstable so that during the restoration of the
masses occupied states of the negative Dirac continuum are partially
transferred into the positive continuum.
The amplitude for this is obtained in a straightforward way within
the In/Out formalism.
A particle with momentum ${\bf p}$ sitting in the negative Dirac sea
is described by the incoming (constituent) wavefunction
$\psi ^{(In)}_{\downarrow } \, \equiv \, \varphi ^{(+)}_{{\bf p}
\downarrow } \, = \, \varphi ^{(c)}_{{\bf p} \downarrow}({\bf x},t)$
(the arrow `$\downarrow $' denotes a Dirac-state in the negative
continuum, while `$\uparrow $'denotes one in the positive continuum).
After the
change of the vacuum structure this wavefunction goes over into
$ \psi ^{(In)}_{\downarrow } \, \rightarrow \,
\varphi ^{(+)}_{{\bf p} \downarrow } \, = \,
\alpha _{{\bf p}} \varphi ^{(b)}_{{\bf p}\downarrow} \, - \,
\beta _{{\bf p}} \varphi ^{(b)}_{{\bf p}\uparrow}$, where states with
the label (b) denote the outgoing (bare) wavefunctions.
The number of produced quarks
is obtained by the projection squared on the appropriate outgoing states
defining the asymptotic particles \cite{MRGr}
and takes here the simple form
\begin{eqnarray}
\label{2}
N_{{\bf p}}^{(Out)} & = &
\langle 0 \, In \mid \hat{b}_{{\bf p}}^{\dagger (Out)} \hat{b}_{{\bf
p}}^{(Out)}
\mid 0 \, In \rangle \nonumber \\
& = &
\beta _{{\bf p}}^{\ast } \beta _{{\bf p}}
\langle 0 \, In \mid \hat{d}_{-{\bf p}}^{(In)} \hat{d}_{-{\bf p}}^{\dagger
(In)}
\mid 0 \, In \rangle
 \, = \, \mid \beta _{{\bf p}} \mid ^2
 \nonumber \\
& = & \frac {1}{2} \left(
1 \, - \, \frac{m_bm_c + {\bf p}^2}{E_{{\bf p}}^{(b)}E_{{\bf p}}^{(c)}}
\right) \, \, \, .
\end{eqnarray}
The number of produced
quarks is depicted in fig. 1. It is found that the spectrum
resembles the distribution of a saturated hot quark-antiquark plasma with a
temperature
between 200 and 300 MeV, although the distribution does not drop off
exponentially, but as a power law for large momenta,
$N_p \approx (m_c - m_b)^2/(4p^2)$. The total number of produced
quarks is hence linearly divergent and should be regulated by a typical
momentum cutoff $\Lambda _C$ as in the NJL model \cite{NJL} which corresponds
to a separation between the low and high momentum degrees of freedom.
For $\Lambda _C \simeq 1$ GeV the total produced quark density is
$n_q \sim 1.5 - 2 $fm$^{-3}$ which again is similar to the quark densities
in a plasma for temperatures
between 200 and 300 MeV. In addition we have also shown in fig. 1 the
situation for a bare quark mass of 150 MeV ($m_c=550$ MeV),
a typical value for the strange quark. Although the production
is suppressed for small momenta, the total integrated quark density is only
minorly affected ($n_s \sim 0.65 - 0.8 $ fm$^{-3}$).

The mechanism for this multiparticle production can be understood as
follows: In the current quark picture, what happens is that the
`hadronic' vacuum state $\mid 0 \, In \rangle $ corresponds to a complex
coherent superposition of current quark states which rapidly go out
of phase.
The time $\Delta t$ needed for the vacuum to restore chiral symmetry
is expected to be
proportional to the inverse of the typical
nonpertubative QCD energy scale
or the chiral symmetry breaking scale
$\Lambda _{QCD}^{-1} \sim \Lambda _C^{-1}
\sim 0.2$ fm/c. A similar estimate is based on the kinetic equilibration
time of the gluons at collider energies predicted to be of the order
$0.3$ fm/c \cite{Shu92,Ge2}.
A small but finite duration $\Delta t$
suppresses the pair production at sufficiently high momenta, because the
wavefunction of the quark resolves the smoothness of the transition.
To quantify this we substitute for the time dependent mass (\ref{1})
\begin{equation}
\label{4}
m^*(t) \,  = \,  \frac{m_c + m_b}{2} - \frac{m_c - m_b}{2} \, \mbox{sgn}(t)
\left( 1-e^{-2|t|/\tau } \right)
\, \, \,
\end{equation}
and propose a nontrivial ansatz for the wavefunction
\begin{equation}
\label{5}
\psi_{{\bf p}} ({\bf x},t) \,  = \,  \frac{1}{\sqrt{V}} \left(
\begin{array}{c}
e^{i\alpha (t)} \cos \varphi (t) \, U \\
e^{-i\alpha (t)} \sin \varphi (t)
\, \frac{\hat{\vec{\sigma}} \cdot \vec{p}}{p} \, U
\end{array}
\right)
\, e^{-\frac{i}{\hbar}(\varepsilon (t)- {\bf p} \cdot {\bf x})} \, \, \, ,
\end{equation}
where $\alpha , \varphi$ and $\varepsilon $ are real function in time.
Inserting the ansatz into
the time dependent Dirac equation yields the following coupled set of
differential
equations
\begin{eqnarray}
\label{6}
\dot{\varepsilon } & = &
\frac{1}{2}p \cos  (2\alpha)  \frac{1}{\sin \varphi \cos \varphi} \nonumber  \\
\hbar \dot{\alpha} & = & - m^*(t) +
p\cos(2\alpha )
\left( \frac{1}{2 \sin \varphi \cos \varphi} -
\frac{\sin \varphi}{ \cos \varphi} \right) \\
\hbar \dot{\varphi} & = & p \sin (2\alpha ) \, \, \, .  \nonumber
\end{eqnarray}
With the appropriate initial condition for the state in the negative
continuum these equations can be integrated numerically and then
projected on the outgoing particle state.
In fig. 2 the spectrum of
particles produced in a `smooth' transition are depicted for various
choices of $\tau $. The total duration of the transition is approximately
$\Delta t \approx 1.5 \,  \tau $. As expected, the particle number at larger
momenta
depends sensitively on the choice of $\tau $. However, for
$\tau < 0.2$ fm/c (and thus $\Delta t \,  < 0.3$ fm/c)
only the high momentum yield is affected.
We conclude that if for dynamical reasons, as estimated above,
the transition from the broken to
the unbroken chiral phase occurs rapidly enough, the change of the underlying
vacuum structure is accompanied by a substantial nonperturbative production of
quarks
and antiquarks.

The additional possibility of gluon decay suggests an interesting
field theoretical problem and can be phrased as follows:
What is the number of quarks
(and antiquarks) produced in a first order decay of already {\em present}
gluons propagating in an external time dependent scalar background field?
We assume now that gluons interact perturbatively with the quarks,
although they are in part responsible for the dynamic quark mass.
The formulation
we have carried out by using real-time Green functions \cite{Ch85,Gr94} is
appropriate for non-equilibrium studies. The number of produced quarks
is contained in the `$<$'-component of the complete one-particle
Green function and is given by the projection on the outgoing state
in the distant future
\begin{eqnarray}
\label{7}
N_{{\bf p  }}^{(Out)} &= &
\langle \Omega ^{(in)} \mid
\hat{U}(-\infty , \infty ) \,
\hat{b}_{{\bf p}}^{\dagger (Out)} \hat{b}_{{\bf p}}^{(Out)}
\hat{U}(\infty , -\infty ) \,
\mid \Omega ^{(in)} \rangle \nonumber \\
& = &
 \lim_{t_1=t_2 \to \infty} \int d^3x_1d^3x_2
\, \left( (-i) \, \varphi _{{\bf p} \uparrow}^{(b) \dagger} (1) \,
G^<(1,2) \, (\gamma_0 \varphi _{{\bf p} \uparrow}^{(b)} (2)) \right)
\, \, \, .
\end{eqnarray}
For our purpose we have to specify the fermionic initial conditions.
For this the zeroth order Green function $G_0$ is determined
by the
full set of wavefunctions defined in respect to the incoming vacuum state
and evolving {\em nonperturbatively}
in time according to the
time dependent scalar background Hamiltonian (\ref{1}), i.e.
\begin{eqnarray}
\label{8}
G_0^< (1,2) & = & +i \sum_{{\bf p}_1}
\varphi _{{\bf p}_1 \downarrow} ^{(+)}(1)
\bar{\varphi} _{{\bf p}_1 \downarrow} ^{(+)}(2)  \nonumber \\
G_0^> (1,2) & = & -i \sum_{{\bf p}_1}
\varphi _{{\bf p}_1 \uparrow} ^{(+)}(1)
\bar{\varphi} _{{\bf p}_1 \uparrow} ^{(+)}(2)  \, \, \, .
\end{eqnarray}
A perturbative expansion is diagrammatically straightforward. The first order
decay is contained in the lowest order self-energy insertion
$\Sigma $ for the quarks, where
the average distribution
$\langle n_{{\bf k}} \rangle $ of the real gluons
enters explicitly.
The number of produced particles can now be calculated
\cite{Gr1}.
By using some general properties of the self energy and the
Green functions,
the change in the quark occupation number induced by the self-energy insertion
reads
\begin{eqnarray}
\label{11}
\Delta N_{{\bf p }}^{(Out)} &= &
2 \cdot \lim_{t_1=t_2 \to \infty} \int d^3x_1d^3x_2
 \, \int d^4x_3d^4x_4 \,
\Re \left\{ (-i) \varphi _{{\bf p} \uparrow}^{(b) \dagger} (1) \,
G_0^{ret} (1,3) \theta (t_3-t_4) \, \right.
\nonumber \\
&&
\hspace*{2cm}
\left. \left[  \Sigma ^{>} (3,4) G_0^< (4,2)
\, - \, \Sigma ^{<} (3,4) G_0^> (4,2) \right]
 \, (\gamma_0 \varphi _{{\bf p} \uparrow}^{(b)} (2)) \right\}
\, \, \, .
\end{eqnarray}
The expression (\ref{11}) for the amount of produced
(or scattered) particles resembles the typical form used in
kinetic transport theories \cite{Ch85,Gr94}; in particular the two terms in
the inner bracket suggest a `Gain' and `Loss' contribution.
Two reasons, however, make further analysis cumbersome. The first, purely
technical one lies in the fact that the Green functions depend on both time
arguments
separately due to the time dependent external field.
The second one stems from the occurrence of a number of infrared singularities.
Besides the pair production the above expression also incorporates the
absorption or emission of gluons for the quarks produced
in lowest order in the vacuum decay (\ref{2}).
A separation of the pair production from these processes is not
possible because of quantum mechanical interference. In the end one has to show
that these singularities cancel each other, as they indeed do \cite{Gr1}.
Finally,
(\ref{11}) can be split into three contributions \cite{Gr1},
\begin{equation}
\label{12}
\Delta N_{{\bf p }}^{(Out)} \, = \,
\Delta N_{{\bf p }}^{`{\cal GAIN}'} \, + \,
\Delta N_{{\bf p }}^{`{\cal LOSS}'} \, + \,
\Delta N_{{\bf p }}^{`{\cal MASSHIFT}'}  \, \, \, ,
\end{equation}
where
the `Gain' and the `Loss' term both are not simply the squared of
some particular amplitude. However, as a numerical investigation shows,
the first is mostly positive while the second is negative.
The interpretation stems from the fact that for example the `Gain' part
contains quarks scattered into the momentum state ${\bf p}$ either by emission
or absorption of a gluon as well as the decay of a gluon into a quark
and antiquark. The exotic process of the simultaneous creation
of a quark-antiquark pair {\em and} a gluon also contributes. The `Loss'
term contains the
opposite processes. The cancellation of the infrared singularities
among both terms is easily understood
within this interpretation.
The designation of a `masshift' part for the last contribution
is based on the known fact that in an equilibrated plasma the quarks
become dressed and dynamically massive by the interaction
with the thermal gluons \cite{We82}. Such a masshift should enter as
a correction to the zeroth order vacuum decay (\ref{2}) \cite{Gr1}.
An ultraviolet divergence
arising from interactions with the perturbative gluonic vacuum
is interpreted as
mass renormalization.
The vacuum terms can explicitly be separated
and should be regarded as the first order vacuum correction to the spontaneous
decay.
However, we are mainly interested in the overall effect due to
the presence of {\em real} gluons.
In the following numerical
study these vacuum contributions are therefore discarded.

The distribution of gluons, at the time where the restoration
of chiral symmetry takes place, is probably far from local thermal or
kinetic equilibrium. In particular
high energetic gluons parallel to the beam axis may be present.
However, for our present purpose it is
convenient and sufficient to parametrize the incoming
gluons by a thermal distribution with a large temperature (in the range
of 700 to 1000 MeV) which includes such high momenta.
The results for a chosen temperature of $T_{g}=700$ MeV are shown
in fig. 3. The QCD coupling constant $\alpha _s=g^2/(4\pi )$ is taken to be
0.3.
For low momenta the quark distribution is large and positive
and exceeds unity by three orders of magnitude. It then quickly turns and
remains negative
up to momenta of 400 MeV. At larger momenta the distribution
resembles a not fully saturated thermal
quark distribution of nearly the same temperature as the gluons.
The distribution,
however, drops faster at still higher momenta which are not shown here.
The low momentum behaviour is obviously unphysical and shows that
pertubation theory
is not applicable here.
If we integrate the calculated distribution over momentum space,
it turns out that the behaviour at low momentum is irrelevant. The major
amount of particles is {\em created} at larger momenta.
An integration up to still only moderate momenta ${\bf p} \simeq  1200$ MeV
explicitly demonstrates that about 12$\times $0.36=4.3 quarks per fm$^{3}$ are
produced by the initiated decay of gluons (integrating up to momenta $\simeq
2500$ MeV
the number changes to 12$\times $ 0.84=10.1 quarks per fm$^{3}$)!
Further numerical investigation shows that the importance of this effect
increases with a higher temperature employed for the gluons. It turns
out that the nearly collinear gluons with high momenta are responsible for
the decay \cite{Gr1}. This suggests that such a phenomenon should be of
particular
importance in the early stages of the gluon evolution when
the chiral symmetry restoration takes
place.
The effect, however, is small if we take a bare mass $m_b$ of 150 MeV.
The distribution is shifted towards higher momenta.
Integrating up to momenta $\simeq 2500$ MeV the number of produced
quarks is evaluated as 0.34 fm$^{-3}$.
Consequently, compared to the light quarks, the
number of strange quarks produced by gluon decay would be strongly suppressed.

To summarize, the occurrence of a rapid chiral restoration of the
vacuum can lead to some particular new and interesting phenomena in
very energetic relativistic heavy ion collisions.
In this letter we investigated
the pair production of quarks within a simple model
illustrating such a dynamical transition.
The presence of real gluons assist the overall production:
During the change of the vacuum structure the direct decay of a gluon into a
quark and
antiquark pair is kinematically allowed. According to our finding
indeed a significant amount of quarks are produced during this process.
Such a phenomenon could give rise to a nonperturbative dynamical creation
of quarks and antiquarks (as well as entropy) in the very early stages
of the heavy ion reaction. It is tempting to speculate that the total
number of produced quarks during such a transition and during the further
evolution suffices so that not only the gluonic degrees of freedom are
saturated, but also the fermionic. The `hot glue' scenario
could turn out to be a hot `quark gluon plasma' after all.

\begin{acknowledgements}

The author wants to thank the Alexander von Humboldt Stiftung
for its support with a Feodor Lynen scholarship. This work was also supported
in part by the U.S. Department of Energy (grant DE-FG05-90ER40592).
The author is thankful for the various and stimulating discussions
with Prof. B. M\"uller. Discussions with A. Sch\"afer and S. Mrowczynski are
appreciated.

\end{acknowledgements}

\begin{figure}
\caption{The spontaneous production of quarks is shown for $m_c=350$ MeV,
$m_b \simeq 0$ MeV (a) and for $m_c = 550$ MeV, $m_b = 150$ MeV (b).
To guide the eye a thermal distribution of massless quarks
is also drawn (for a temperature of $T=200$ MeV (c) and $T=300$ MeV
(d)).}
\label{fig1}
\end{figure}

\begin{figure}
\caption{The production of quarks is shown for a `smooth' transition in
time of the vacuum (compare text). The parameters given in the figure are
to be read from the left to the right. $m_c=350$ MeV and $m_b$ is taken
to be nearly zero.}
\label{fig2}
\end{figure}

\begin{figure}
\caption{The first order corrections on the particle distribution
due to the presence of real gluons are depicted. The gluons are taken to
be distributed with a temperature of 700 MeV.
(a) corresponds to the parameter set $m_c=350$ MeV, $m_b = 10$ MeV,
(b) to the same as in fig. 1. For a comparison a thermal distribution
of massless quarks with a temperature of 700 MeV is also drawn (c).}
\label{fig3}
\end{figure}


\begin{references}
\bibitem{Gr1}
The extensive analytic calculation
will be presented in another publication, where
the explicit
cancellation of all occurring IR-singularities will be discussed in detail.
\end{references}
\end{document}